\documentclass[12pt]{article}
\usepackage{graphics}

\baselineskip

\begin{document}
\vspace*{-0.8cm}
\begin{flushright}
{\large cond-mat/YYYY}\\ {April-2000}\\
\end{flushright}

\newcommand{\figuno}{
\begin{figure}[tb]
\begin{center}
\includegraphics{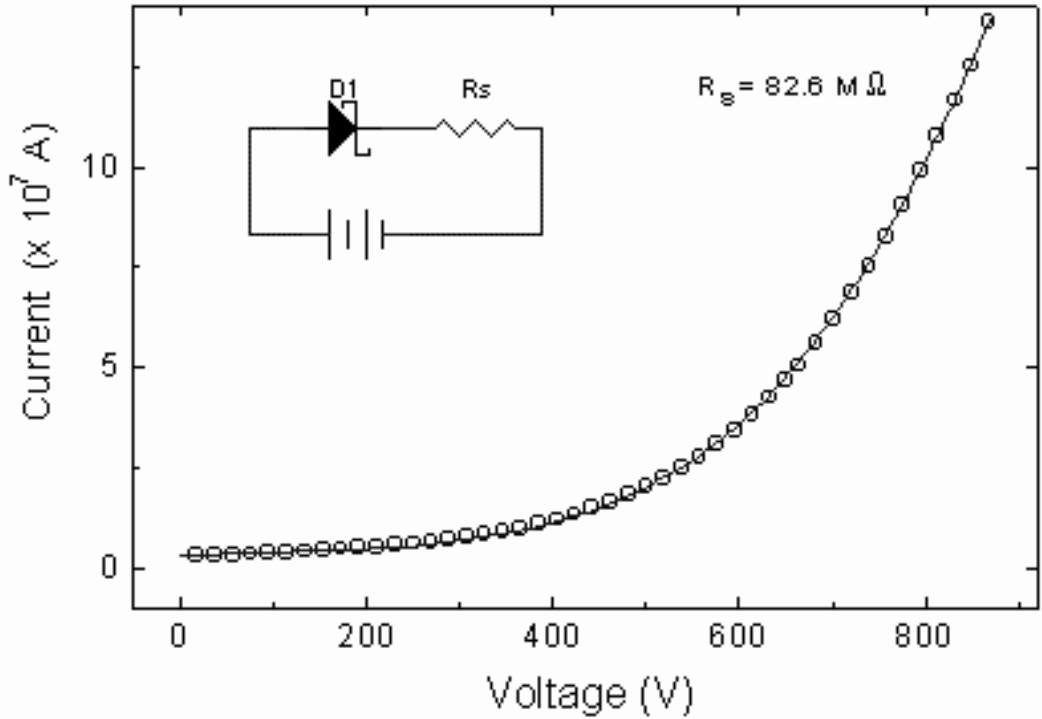}
\caption{Direct current I-V characteristic at 313 K for an
${Al}_2 O_3$:$Mg$ crystal
containing $[Mg]^{0}$ centers. The solid line represents the best fit
of experimental points to a directly biased barrier with a series resistance.}
\label{fig:uno}
\end{center}
\end{figure}
}

\newcommand{\figdos}{
\begin{figure}[tb]
\begin{center}
\includegraphics{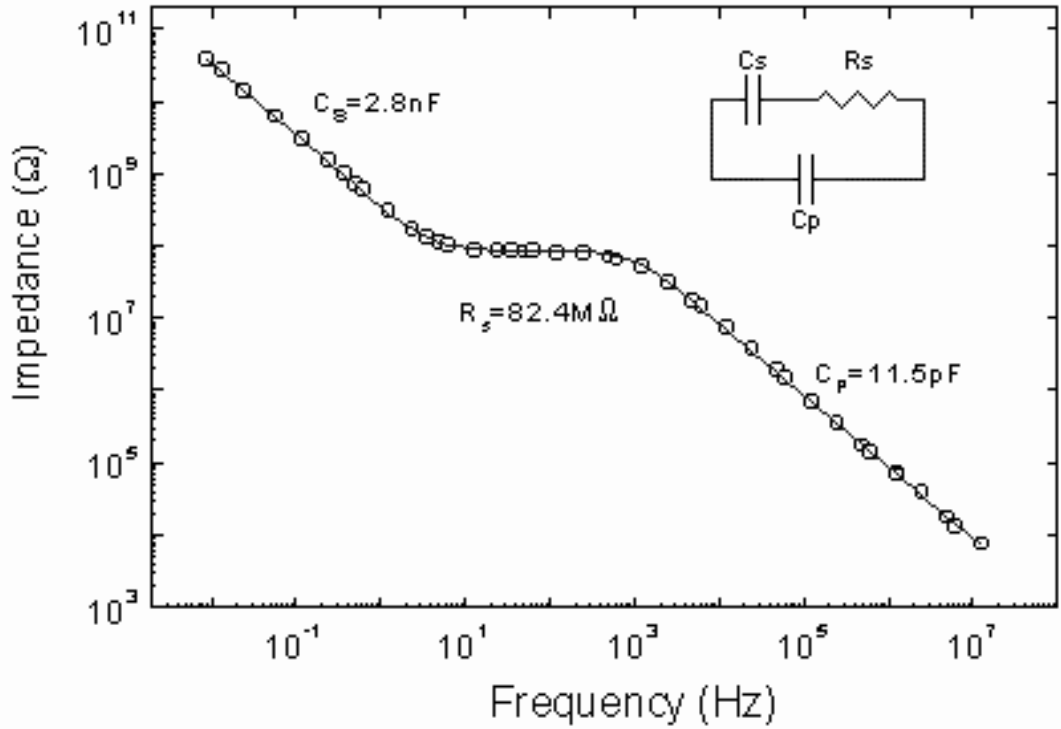}
\caption{Log-log of impedance versus frequency at 313 K for an ${Al}_2 O_3$:$Mg$ crystal
containing $[Mg]^{0}$ centers. The solid line represents the best fit of the
experimental points to the equivalent circuit.}
\label{fig:dos}
\end{center}
\end{figure}
}

\newcommand{\figtres}{
\begin{figure}[tb]
\begin{center}
\includegraphics{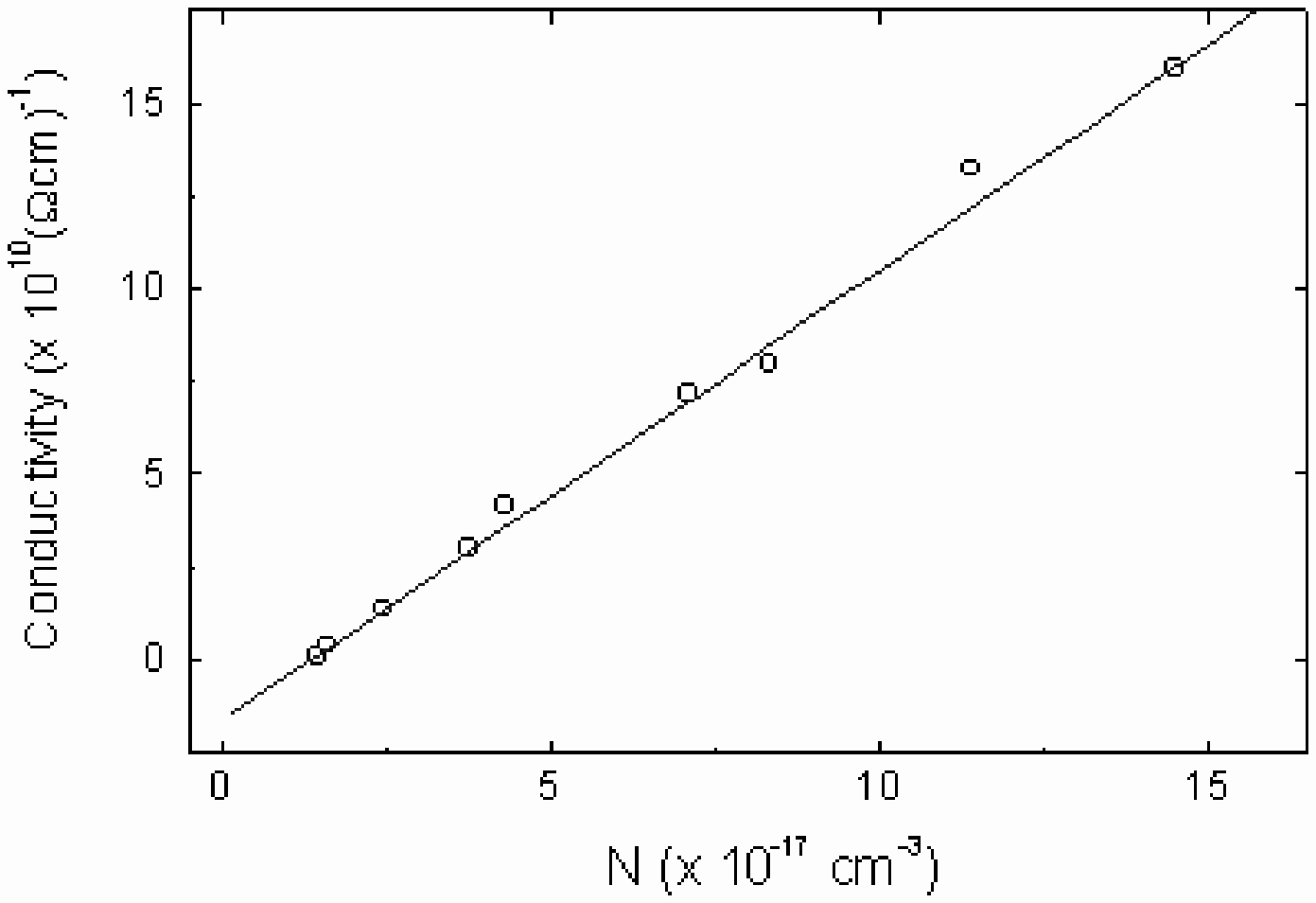}
\caption{Conductivity versus concentration of $[Mg]^{0}$ centers at  T=313 K. The
concentration values were determined from the peak of the optical absorption
band at 2.5 eV associated with the $[Mg]^{0}$ centers~\cite{sie}.}
\label{fig:tres}
\end{center}
\end{figure}
}

\begin{center}
\begin{large}
\begin{bf} ELECTRICAL CONDUCTIVITY IN \\ MAGNESIUM-DOPED {\bf ${Al}_2 O_3$} CRYSTAL
AT MODERATE TEMPERATURES.\\
\end{bf}
\end{large}
\vspace{0.4cm} M. Tard{\'\i}o $^{\star}$, R. Ram{\'\i}rez, R.
Gonz{\'a}lez\\ \vspace{0.05cm} {\em  Departamento de F{\'\i}sica\\
  Escuela Polit{\'e}cnica Superior\\
Universidad Carlos III de Madrid\\ Avda. de la Universidad 30,
Legan{\'e}s, Madrid,\ \ Spain} \\ \vspace{0.05cm} Y. Chen\\
\vspace{0.06cm} {\em  Division of Materials Sciences\\
 U. S. Department of Energy, Germantown\\
Maryland 20874-1290, USA}\\
\vspace{0.05cm}
and\\
\vspace{0.1cm}
M.R. Kokta\\
{\em BICRON Crystal Products, Washougal\\
Washington 96871, USA}
\end{center}

\begin{center}
{\bf ABSTRACT}
\end{center}
\begin{quotation}
\noindent
AC and DC electrical  measurements between $273$ and $800$ K were used to
characterize the electrical  conductivity of ${Al}_2 O_3$:$Mg$ single crystals
containing $[Mg]^{0}$ center. At  low fields contacts are blocking. At high fields,
electrical current flows steadily through the sample and the I-V characteristic
corresponds to a directly biased barrier whit a series resistance (bulk
resistance). AC measurements yield values for the junction capacitance
as well as for the sample resistance, and provide perfectly reproducible
conductivity values. The conductivity varies linearly whit the $[Mg]^{0}$
concentration and a thermal activation energy of $0.68$ eV was obtained,
which agrees very well with the activation energy previously reported for
motion of free holes.
\end{quotation}
\vspace{0.05cm}
\vspace{0.4cm}
${\star}$ {\em e-mail: tardio@alum.fis.uc3m.es}\\

\newpage

\section{INTRODUCTION}

$MgO$ crystals  containing $[Li]^{0}$ centers (substitutional
${Li}^{+}$ ions, each atten\-ded by a hole), have been shown to be
p-type semiconducting, with an acceptor level of $0.7$
eV~\cite{uno}-\cite{ci}. By analogy, substitutional ${Mg}^{2+}$ in
${Al}_2 O_3$  is also atten\-ded by hole~\cite{seis,sie} and
expected to serve as a p-type semiconductor. Both $MgO$: $Li$ and
${Al}_2 O_3$:$Mg$ systems have potential as high-temperature
p-type semiconductors. However whereas the former is brittle and
will have limited application, the latter is expected to have
better mechanical integrity.  In  the present study, AC and DC
electrical measurements were used to charac\-te\-ri\-ze the
electrical conductivity of ${Al}_2 O_3$:$Mg$ single crystal
containing $[Mg]^{0}$ center after oxidation at elevated
temperatures.\\

\section{EXPERIMENTAL PROCEDURE}
${Al}_2 O_3$:$Mg$ crystal were grown by the Czochralski method.
Atomic Emi\-ssion Spectrometry analyses indicated that the
magnesium concentration was $25$ppm. Samples were polished to
optical transparency. The c axis was pa\-ra\-llel to the broad
face of the sample. Electrodes were made by sputtering several
metals with different work functions ($Mg$, $Al$ and $Pt$) onto
the samples surfaces, and the electrical field was applied
perpendicular to the c axis. The same response was observed
regardless of contact electrode materials. Optical absorption
measurements were made whit Perkin Elmer Lambda $19$
spectrophotometer.\\

\section{RESULTS AND DISCUSSION}
Single crystal of ${Al}_2 O_3$:$Mg$ are colorless after heating at
$T > 950$ K in a
reducing atmosphere. However they become gray-purple after oxidation at
$T > 1200$ K~\cite{sie,ocho}. This coloration is due a broad asymmetric
optical adsorption
band, centered at $\approx 2.5$ eV ($496$ nm), and has been attributed to
${Mg}^{2+}$ cations,
each with a trapped hole localized on one of the six NN oxygen ions. These
defects are paramagnetic and are referred to as $[Mg]^{0}$ centers~\cite{tres,cua}.

Electrical conductivity measurements between $273-800$ K were made in
${Al}_2 O_3$:$Mg$ crystal containing $[Mg]^{0}$ centers. DC electrical measurements at low
fields revealed blocking contacts. However, at high fields, the reverse bias
characteristic is that of a "soft" barrier, and the I-V characteristic of the
sample corresponds to a directly biased barrier with a series resistance, Rs,
(bulk resistance), regardless of the polarity of the applied voltage (Fig.1).

\figuno

The results of low voltage AC measurements reinforce the DC interpretation
and yield values for the junction capacitance as well as for the sample
resistance. Fig. 2, shows a log-log plot of impedance the applied voltage
frequency for a low voltage ($\approx 1$V). Three frequency regions are
observed, consistent whit a series combination of a capacitance ($Cs$),
with a resistance
($Rs$), and  both  connected in parallel with a capacitance ($Cp$).
In Fig.2, the
corresponding impedances of  $Cs$, $Rs$, $Cp$ dominate the low,
intermediate and high
frequency region, respectively. The resulting value for  $Cp$ agrees with the
capacity of a parallel-plate condenser with the ${Al}_2 O_3$ dielectric
constant. $Rs$ is in good agreement with the value derived from
${Al}_2 O_3$ I-V characteristic.

\figdos

Lastly a $Cs$ value of the order of a nF seem to be reasonable for
a sample with a cross section of $25 mm^{2}$.

AC measurements  provide perfectly reproducible conductivity
values. At a fixes temperature, the conductivity varies linearly
with the $[Mg]^{0}$ concentration (Fig. 3). From the temperature
dependence of the conductivity, a thermal activation energy of
$0.68$ eV was obtained, which agrees very well with the activation
energy for the motion of free holes as large polarons~\cite{sie}.

\newpage
\figtres

\section*{Acknowledgments.}
Research at the University Carlos III was supported by the CICYT of Spain. The
research of Y.C. is an outgrowth of past investigation performed at the Oak
Ridge National Laboratory. Chemical analyses were performed at the Centro
de Espectrometr{\'\i}a At{\'o}mica de la Uiversidad Complutense de Madrid.

\end{document}